\documentclass[%
showkeys,
preprint,
amsmath,amssymb,
pra,
floatfix,
longbibliography
]{revtex4-2}
\usepackage{subfig}
\usepackage{comment}
\usepackage{graphicx}% Include figure files
\usepackage{dcolumn}% Align table columns on decimal point
\usepackage{bm}% bold math
\usepackage{float}
\usepackage{stackengine}
\usepackage{xcolor}
\newcommand{\ket}[1]{\left\vert#1\right\rangle}

\begin{document}

	
	
	\title{Impact of dephased entangled states and varying measurement orientations on the reliability of cryptographic keys generated via the quantum protocol E91: A quantum simulation approach}
	
	
\author{Adri{\'a}n F. Hern{\'a}ndez-Borda}
%\email{adrian.hernandez@uptc.edu.co}
\affiliation{
	Grupo QUCIT, Escuela de F{\'i}sica, Universidad Pedag{\'o}gica y Tecnol{\'o}gica de Colombia, Tunja 150003, Colombia}%
%\affiliation{%
	%Grupo de F{\'i}sica Te{\'o}rica y Computacional, Escuela de F{\'i}sica, Universidad Pedag{\'o}gica y Tecnol{\'o}gica de %Colombia,Tunja 150003, Colombia}

\author{María P. Rojas-Sepúlveda}
%\email{adrian.hernandez@uptc.edu.co}
\affiliation{
	Grupo QUCIT, Escuela de F{\'i}sica, Universidad Pedag{\'o}gica y Tecnol{\'o}gica de Colombia, Tunja 150003, Colombia}%
%\affiliation{%
	%Grupo de F{\'i}sica Te{\'o}rica y Computacional, Escuela de F{\'i}sica, Universidad Pedag{\'o}gica y Tecnol{\'o}gica de %Colombia,Tunja 150003, Colombia}

\author{Hanz Y. Ram{\'i}rez-G{\'o}mez}%
\email{hanz.ramirez@uptc.edu.co}
\affiliation{
	Grupo QUCIT, Escuela de F{\'i}sica, Universidad Pedag{\'o}gica y Tecnol{\'o}gica de Colombia, Tunja 150003, Colombia}% v
%\affiliation{%
	%Grupo de F{\'i}sica Te{\'o}rica y Computacional, Escuela de F{\'i}sica, Universidad Pedag{\'o}gica y Tecnol{\'o}gica de %Colombia,Tunja 150003, Colombia}


	%% Abstract
	\begin{abstract}
		Photons and optical circuits are among the most promising platforms for implementation of quantum technologies, because of its potential use in quantum computing and long-distance quantum communication, including quantum cryptography.
		
		One of the main requirements to achieve reliable quantum communications are on-demand sources of highly entangled photon pairs, and semiconductor quantum dots have emerged as prominent candidates to satisfy the necessary conditions of brightness and entanglement fidelity.
		
		However, in most cases the biexciton-exciton-vacuum cascade produces a pair of maximally polarization-entangled photons with a dephasing, due to a non-negligible exciton fine structure splitting in the emitting nanostructure.
				
		This work focuses on the performance of the E91 quantum key distribution protocol under the variation of two elements: first, the phase in the input state when the protocol is implemented using entangled photons generated via the radiative cascade, and second, the relative directions of the polarization analyzers. 
				
		We use a quantum computational approach by means of the IBM's API Qiskit to simulate the optical implementation of the studied cryptographic protocol and thus to validate analytical expressions derived for the secret key rate and the Bell's parameter, given as functions of the input state's phase and of the polarization measurement angles.   
				
		Our results show that the performance of the quantum transmission is highly impacted by the product between the exciton lifetime and the quantum dot's fine structure splitting and that such an impact may be modulated through the orientation of the polarizers. Under some specific conditions, the studied E91 protocol is shown to turn into the BBM92 protocol, to which the results can be extended. 
				
		These findings provide important insight for the scalable implementation of quantum key distribution protocols with realistic entanglement sources. 
		
		Furthermore, this study constitutes an illustrative example of how quantum computation can be used as a powerful tool for simulating physical processes whose experimental realization can be substituted by short algorithms run on quantum software.   
		
		
	\end{abstract}

\maketitle
\newpage
	
%	\begin{graphicalabstract}
%		%\includegraphics{grabs}
%	\end{graphicalabstract}
	
%	\begin{highlights}
%		\item Research highlight 1
%		\item Research highlight 2
%	\end{highlights}
	
	%% Keywords
%	\begin{keyword}
%		Quantum simulation \sep Quantum Dots \sep Fine Structure Splitting \sep Quantum Key Distribution \sep Entangled-photon Sources
%		% keywords here, in the form: keyword \sep keyword
%		
%		%% PACS codes here, in the form: \PACS code \sep code
%		
%		%% MSC codes here, in the form: \MSC code \sep code
%		%% or \MSC[2008] code \sep code (2000 is the default)
%		
%	\end{keyword}
%	
%\end{frontmatter}
	
	%\tableofcontents
\section{Introduction}\label{sec:1_introduction}
	%--------------------------------------------------------------------------------------------------------
	Quantum key distribution (QKD) is a private-key cryptographic model that exploits the principles of quantum mechanics such as superposition, entanglement, and collapse of quantum states, to transmit quantum bits that conform a cryptographic key to encrypt and decrypt information \cite{bennett1983quantum,ekert1991quantum,nielsen2010quantum,ilic2007ekert}.
	%--------------------------------------------------------------------------------------------------------
	Realization of this type of cryptographic scheme has been achieved mainly with the use of optical quantum entangled states, what has promoted the photonic implementation as a realistic option for quantum communication and quantum internet \cite{bennett1992experimental,ekert1992practical,peng2007experimental,yin2016measurement,wengerowsky2018entanglement,yin2020entanglement}.
	%--------------------------------------------------------------------------------------------------------
	Hence, the development of optimized entangled-photon sources has become a topic of intense research toward the progress of light-based quantum technologies \cite{wang2019demand,meng2024deterministic,chen2024wavelength,weissflog2024tunable}.
	%--------------------------------------------------------------------------------------------------------
	
	%--------------------------------------------------------------------------------------------------------
	Quantum entangled states were widely discussed and subject of controversy during the second part of the XX century since their introduction by Einstein, Podolsky, and Rosen \cite{einstein1935can}.
	%--------------------------------------------------------------------------------------------------------
	Arguably the main contribution toward resolving the dispute on the reality of those perplexing states was provided by J. Bell, who mathematically proved that any hidden variable theory is incompatible with the statistical predictions of quantum mechanics. Its result is known as the Bell's Theorem \cite{bell1964einstein}.
	%--------------------------------------------------------------------------------------------------------
	Later on, other authors recreate their own versions of that theorem, raising a set of expression named Bell inequalities. Particularly, Clauser et al. in reference \cite{clauser1969proposed}, develop a theorem known as the CHSH inequality.
	%--------------------------------------------------------------------------------------------------------
	This inequality has been experimentally verified via photon-based experiments in which, first, a radiative cascade decay in calcium atoms, and later, spontaneous parametric down conversion (SPDC) in nonlinear crystals, were used as entangled-photon sources \cite{aspect1981experimental,aspect1982experimental,tittel1998violation,fujiwara2014modified}.
	%--------------------------------------------------------------------------------------------------------
	
	%--------------------------------------------------------------------------------------------------------
	The so-called E91 protocol was the first quantum entanglement-based QKD protocol, devised by A. K. Ekert in the early 90's \cite{ekert1991quantum,wang2021chip}.
	%--------------------------------------------------------------------------------------------------------
	In the original version of the protocol, the author proposed the use of $1/2$-spin particles prepared in a Bell's state. Nevertheless, at the last part of that seminal work, he mentioned that an optimal realization could be based on correlated photon states.
	%--------------------------------------------------------------------------------------------------------
	The protocol includes a mechanism for validating the security of the transmitted key based on measurements of the Bell's theorem, which allows to rule out eavesdropping. That verification mechanism is normally implemented in terms of the CHSH inequality. 
	%--------------------------------------------------------------------------------------------------------
	This QKD scheme was first effectively implemented by using polarization-entangled photon states produced via SPDC. Afterwards, entangled photons obtained from semiconductor quantum dots (QDs) were used, achieving long-distance transmission, although the yielded key was found below what could be achieved with a SPDC source   \cite{ekert1992practical,ling2008experimental,fujiwara2009performance,fujiwara2014modified,dzurnak2015quantum,basso2021quantum}.
	%--------------------------------------------------------------------------------------------------------
	
	%--------------------------------------------------------------------------------------------------------
	SPDC has been so far the most commonly employed mechanism for entanglement generation, because of the highly entangled photons obtained by this method \cite{tittel2001experimental}.
	%--------------------------------------------------------------------------------------------------------
	However, its production of photon pairs is random and then unsuitable for reliable quantum communication or other applications that require an on-demand source of entangled states.
	%--------------------------------------------------------------------------------------------------------
	In this scenario, QDs have appeared as promising candidates for optimal on-demand entanglement generation.
	%--------------------------------------------------------------------------------------------------------
	In fact, there have been several successful implementations of long-distance QKD with QDs as entangled-photon sources \cite{benson2000regulated,yin2016measurement,yin2017satellite,basso2021quantum,morrison2023single,yu2023telecom,zahidy2024quantum,miloshevsky2024cmos}.
	%--------------------------------------------------------------------------------------------------------
	Nonetheless, QD-produced entangled states are frequently dephased with respect to the ideal Bell states for which the QKD protocols are usually designed.
	%--------------------------------------------------------------------------------------------------------
	Such a dephasing is underlaid by the so-called electron-hole exchange, that causes the exciton fine structure splitting (FSS) in strongly confined nanostructures    
	\cite{own-PRL,winik2017demand,own-con-mat,pennacchietti2024oscillating,own-PSS-b}.
	%--------------------------------------------------------------------------------------------------------
	
	%--------------------------------------------------------------------------------------------------------
	In this work we study the effects of the FSS-driven dephasing in the entangled input state and those of the relative orientation between measurement axes on the performance of the E91 QKD protocol. We address the problem with the aid of quantum computation, which allows us to replace the corresponding laborious optical experiment with a succinct implementation in the IBM's API Qiskit.
	%--------------------------------------------------------------------------------------------------------
	In the first part, we introduce the dephasing effects of the FSS on the entangled states produced in QDs. Afterward, we explore the influence of both, the FSS-driven dephasing and the orientation of the detection axes in the execution of the QKD protocol E91, and derive the corresponding analytical expressions. Finally, we validate our model by means of the quantum computing implementation and discuss the impact of the analyzed variables on the performance of the key distribution process.
	%--------------------------------------------------------------------------------------------------------
	
	%--------------------------------------------------------------------------------------------------------
\section{Sources of entangled photon pairs for QKD}\label{sec:2_FSS on biexciton}
	%--------------------------------------------------------------------------------------------------------
	Many of the fundamental experiments that allowed to verify the physical reality behind entanglement, as well as most of the QKD experiments carried out to date, have used SPDC to produce quantum correlated photons \cite{ekert1992practical,bennett1992experimental,naik2000entangled,peng2007experimental,chang2020observation}.
	%------------------------------------------------------
	This mechanism permits the creation of a pair of lower-energy daughter photons originated from a higher-energy pump photon, that interacts with a transparent non-linear crystal inside which it is randomly split into an polarization entangled pair \cite{hong1985theory}.
	%------------------------------------------------------
	Because SPDC is not an on-demand process, its use for implementations of scaled quantum communication is dubious. The probability of obtaining entangled pairs given an incident photon, is described by a Poissonian distribution. Hence, probabilistically none, one or several pairs of entangled photons may be produced, being none much more likely under normal conditions than the other components. Thus, the efficiency of this entanglement-generation method is very low and its potential for applications in emerging quantum technologies clearly limited \cite{guilbert2014enhancing,muller2014demand,couteau2018spontaneous}.
	%------------------------------------------------------
	
	%------------------------------------------------------
	Instead of a stochastic source, quantum dots have been proposed as a deterministic on-demand source of highly entangled photon pairs by exploiting the recombination of the biexciton state ($\ket{XX}$). This process yields two polarization-entangled photons that, in contrast to SPDC, can be successfully applied in quantum communications \cite{benson2000regulated,winik2017demand,huber2018strain,own-sci-rep,liu2019solid}.
	%------------------------------------------------------
	Nevertheless, the coherence of the entangled polarization state may be altered as a result of the interactions between the produced photons and its environment, such as recapture or depolarization by defects; or as consequence of the exciton fine structure associated to intrinsic characteristics of the emitting QD. In particular, this latter effect has been widely studied and identified as the main challenge toward reliable generation of entangled states from QDs \cite{own-old-PSS,vajner2024demand}.
	 
	The dephasing introduced by the FSS into the entangled output state, is known to directly depend on that energy splitting between the exciton states ($\ket{X,-1}$ and $\ket{X,1}$) \cite{young2009bell,huber2018strain,liu2019solid,pennacchietti2024oscillating}.
	%------------------------------------------------------ 
	
	%------------------------------------------------------
\subsection{Entangled photon pairs from radiative cascades in QDs}
	%------------------------------------------------------ 
	
\begin{figure}[H]
		\centering
		\includegraphics[width=.5\columnwidth]{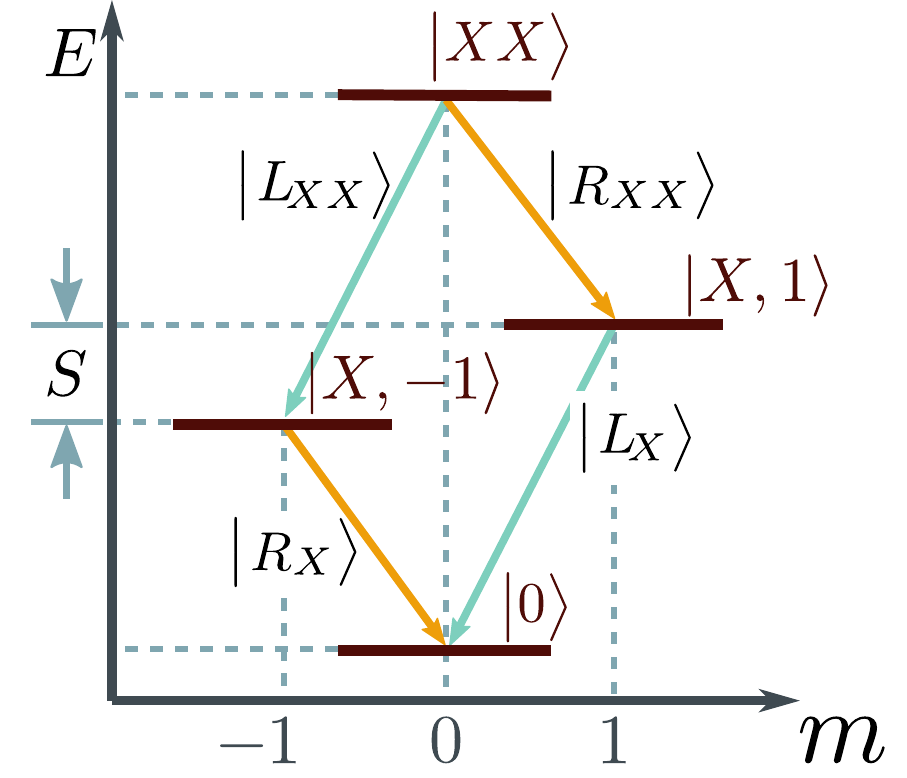}
		\caption{Radiative cascade $\ket{XX} \to \ket{X} \to \ket{0}$ in a quantum dot. There are two different decay routes, one through state $\ket{X,-1}$ ($m=0 \rightarrow m=-1 \rightarrow m=0$) and the other one through state $\ket{X,1}$ ($m=0 \rightarrow m=1 \rightarrow m=0$). The cyan (orange) line represents emission of a right (left) circularly polarized photon.}
		\label{fig:decay_modes}
	\end{figure}

	%------------------------------------------------------
	The process starts with excitation of the neutral biexciton state $\ket{XX}$, that later decays into one of the single exciton states $\ket{X}$ by one of two possible decay routes (either to $\ket{X,-1}$ and $\ket{X,1}$).
	%------------------------------------------------------
	Through the first pathway, a photon with right circular polarization $\ket{R_{XX}}$ is emitted and the system's $z$-component of angular momentum passes from $m=0$ to $m=-1$ (decaying to the state $\ket{X,-1}$).
	%------------------------------------------------------
	Contrarily, through the second pathway, the emitted photon has left circular polarization $\ket{L_{XX}}$ and the system passes from $m=0$ to $m=1$ (decaying to the state $\ket{X,1}$).
	%------------------------------------------------------
	Afterwards, the corresponding exciton state decays into the ground state ($\ket{0}$) emitting a photon with opposite polarization respect to the previously emitted, and the system returns to $z$-component of total angular momentum $m=0$ \cite{winik2017demand,ozfidan2015theory}.
	%------------------------------------------------------
	%For $m=-1$ a $\ket{H_{X}}$ photon is emitted, whereas for $m=1$, a $\ket{V_{X}}$ photon is emitted.
	%%------------------------------------------------------
	%This restores de initial value of $m$ because of the conservation of angular momentum \cite{winik2017demand,ozfidan2015theory}.
	%------------------------------------------------------
	This radiative cascade and its two pathways are depicted in figure \ref{fig:decay_modes}, where cyan (orange) represents emission of a right (left) circularly polarized photon and $S$ stands for the FSS energy.
	%------------------------------------------------------
	%------------------------------------------------------
	
	%------------------------------------------------------
	In the case $S=0$, the two decay pathways are indistinguishable. However, if $S \ne 0$ the degeneracy of exciton states with different angular momentum is lifted.
	%------------------------------------------------------
	This makes the decay routes distinguishable and the two-photon entangled state becomes
	%------------------------------------------------------
	\begin{equation}\label{eqn:singlet_state_dephased}
		\vert\psi\rangle = \dfrac{1}{\sqrt{2}}\left[\vert L_{XX} R_{X} \rangle + e^{-i\theta_\text{FSS}} \vert R_{XX} L_{X} \rangle
		\right],
	\end{equation}
	%------------------------------------------------------
	where the relative phase $\theta_\text{FSS}=S\tau/\hbar$, depends on the splitting energy $S$ and on the time between the first and second electron-hole recombination $\tau$ (neutral exciton radiative lifetime). \cite{hernandez2023effects,winik2017demand,bauch2024demand}.
	%------------------------------------------------------
	
	This state is maximally entangled disregarding the value of $\theta_\text{FSS}$, because the relative phase does not affect the Von Neumann entropy of the system. Nevertheless, such a dephasing creates an oscillation between Bell states along time. In other words, the electron-hole exchange induces via the $FSS$, a local unitary transformation over the Bell singlet state in the Poincare's sphere that affects its stationary character   \cite{fognini2019dephasing,pennacchietti2024oscillating,kuroda2013symmetric}.
	
	The state in equation \ref{eqn:singlet_state_dephased} can be rewritten in terms of linear horizontal (H) and vertical (V) polarizations, according to
	
	\begin{equation}\label{eqn:singlet_state_dephased-linear}
		\vert\psi\rangle\,=\, \dfrac{1}{\sqrt{2}}\left[\vert H_{XX} H_{X} \rangle + e^{-i\theta_\text{FSS}} \vert V_{XX} V_{X} \rangle
		\right].
	\end{equation}
	
	Tuning of the FSS in QDs has been a field of intense research, since the related dephasing produces time dependent oscillations in the fidelity, hurting the possibility of determining the degree of entanglement of the emitted states \cite{review-entanglement-source}.
	
	Different ways of reducing the $S$ energy in QDs have been proposed, aiming to reshape the wave function of the confined carriers, to compensate the asymmetries that strengthen the magnitude of the  electron-hole exchange interaction \cite{RAMIREZ20101155,PhysRevB.81.245324,DeGreve2011,own-sci-rep,PhysRevLett.129.193604}.
	
	Although successful QD tuning has been realized using either piezoelectric substrates or stark effect with external electric fields  \cite{trotta2015energy,huber2018strain,muller2009creating,dusanowski2022all,chen2024wavelength}, still most grown QD samples exhibit non-vanishing FSS due to the inherent lack of rotational symmetry associated to both, the microscopic crystalline structures and the non-perfectly axial dot shapes.
	
	%--------------------------------------------------------------------------------------------------------
	
	%--------------------------------------------------------------------------------------------------------
\section{QKD protocol E91}\label{sec:3_E91 Protocol}
	%--------------------------------------------------------------------------------------------------------
	The original E91 protocol aims the communication of a private key between two distant subjects, Alice and Bob, encoded on the spin of a pair of entangled $1/2$-spin particles prepared on a singlet state. However, Ekert ended suggesting that an optical implementation could be more suitable \cite{ekert1991quantum}.
	%-----------------------------------------------------------------------------------------------------
	Such implementation with photons was actually realized around a year later \cite{ekert1992practical}. The protocol begins with the preparation of a  polarization-entangled Bell state, then one photon is sent to Alice and the other one to Bob. Each participant is expected to measure his/her corresponding particle with adjustable polarizers.
	%-----------------------------------------------------------------------------------------------------
	Each apparatus (Alice's and Bob's) can be set in three possible directions that are part of a group of four preset orientations, defined by its angle with respect to vertical axis. These orientations are labeled $\phi_{i}$, $i=0,1,2,3$, and their angles in the photonic version of the original protocol are defined according to  $\phi_{\ell}=\ell\pi/8$ for $\ell=0,1,2,3$, as illustrated in figure \ref{fig:ekerts_directions}(a). %with $\alpha=\beta=\pi/4$
	%-----------------------------------------------------------------------------------------------------
	
	Afterwards, Alice  randomly selects one of the first three directions and registers her choice ($\phi_{i_A}$). Then, measures the particle and records either $-1$ or $1$, depending on the result of her measurement. In turn, Bob selects one of the last three directions, and also records his chosen orientation ($\phi_{i_B}$) and the corresponding measurement.
	
	In this form, the second and third directions among the set of four ($\phi_{1}=\pi/8$ and $\phi_{2}=\pi/4$), are part of the Alice's and Bob's possibilities, and then, the ones in which there can be coincidence.
	%--------------------------------------------------------------------------------------------------------
	Once the transmission has concluded, each participant shares on a public channel his/her list with the selected orientations for each event.
	%-----------------------------------------------------------------------------------------------------
	Each participant analyzes the other's list and compares it with his/her own to separate his/her registered measurements in two groups: the first contains the measurements for which Alice and Bob chose the same direction, and the second includes the ones taken along mismatched orientations.
	%--------------------------------------------------------------------------------------------------------
	The cryptographic key is the string of results in the first group, while the second group of measurements is employed to validate the security of the key distribution via Bell's test on the CHSH inequality.
	%--------------------------------------------------------------------------------------------------------
	The particular angles defined as multiples of $\pi/8$ were intentionally picked to maximize the quantity 
	
	\begin{equation}
		\label{eqn:CHSH}
		CR= E(\phi_{0},\phi_{1}) + E(\phi_{2},\phi_{3}) + E(\phi_{0},\phi_{3}) - E(\phi_{2},\phi_{1}),
	\end{equation}
	
	given in terms of the correlation amplitudes  
	
	\begin{equation}
		\label{eqn:coefficient_correlation}
		\begin{split}
			E(\phi_{a},\phi_{b}) =
			-\cos\left[2(\phi_{a}-\phi_{b})\right].
		\end{split}
	\end{equation}
	
	%where $C_\phi=\cos(\phi)$.
	
	Such a quantity ($CR$) is used for the Bell's test, carried out in the validation stage of the protocol when eavesdropping is considered. For those specifically chosen angles $CR = -2\sqrt{2}$ \cite{ekert1991quantum,clauser1969proposed,ekert1992practical,kuroda2013symmetric}.
	%--------------------------------------------------------------------------------------------------------
	
\subsection{Modifications to the E91 protocol and analytical results}
	%--------------------------------------------------------------------------------------------------------
	We now consider a polarization-entangled state of photons produced via a QD radiative cascade, as described in section \ref{sec:2_FSS on biexciton}. Additionally, we include a generalization  by defining the detection orientations $\phi_{1}$ and $\phi_{2}$ in terms of variable angles.
	
\begin{figure}[H]
		\centering
		\includegraphics[width=.5\columnwidth]{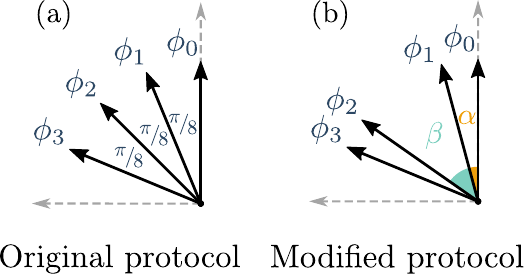}
		\caption{Angles for the optical implementation of the E91 protocol. (a) Fixed measurement directions in the originally proposed version of the protocol. (b) Variable measurement directions in the modified version of the protocol.}
		\label{fig:ekerts_directions}
	\end{figure}

	Thus, we introduce the parameter $\alpha$, as the angle between the vertical axis and the second orientation in the set of four ($\phi_{1}$). Similarly, we define the parameter $\beta$ as the angle between the two coincident directions ($\phi_{2} - \phi_{1}$). These parameters are highlighted respectively in cyan and orange, in figure \ref{fig:ekerts_directions}(b).  
	
	%--------------------------------------------------------------------------------------------------------
	%The first step is focusing on the directions $\phi_{1}$ and $\phi_{2}$, involved in the transmission of the key i.e., the ones that, according to quantum mechanics, can produce a fully determined anticorrelation between the polarization states.
	%--------------------------------------------------------------------------------------------------------
	
	%--------------------------------------------------------------------------------------------------------
	%The second parameter, $\beta$, is the angle between the two coincident directions $\phi_{1}$ and $\phi_{2}$, indicated with yellow in figure \ref{fig:modified_E91}
	%--------------------------------------------------------------------------------------------------------
	
	%--------------------------------------------------------------------------------------------------------
	If we consider two bases, i.e. the vertical-horizontal and the $\phi$-rotated-$\phi$-antirotated linear polarizations, for a Bell state of polarization-entangled photons $\dfrac{1}{\sqrt{2}}\left[\vert H_{XX} H_{X} \rangle +
	\vert V_{XX} V_{X} \rangle
	\right]$ ($\theta_\text{FSS} = 0$), measurements on both photons are certainly correlated if they are obtained in the same basis (along the same orientation), disregarding the angle $\phi$ between those bases \cite{grynberg2010introduction}.
	
	However, if the dephasing associated to the FSS is included ($\theta_\text{FSS}\ne 0$), the probability of correlation will depend on the rotation angle, according to
	
	\begin{equation}\label{eqn:anticorrelation}
		P_{\pm\pm}(\phi,\theta_\text{FSS}) = \Big[C_{\phi}^{4} + S_{\phi}^{4} +2S_{\phi}^{2} C_{\phi}^{2} C_{\theta_\text{FSS}}\Big],
	\end{equation}
	
	where $S_{\phi}=\sin(\phi)$ and $C_\phi=\cos(\phi)$. $P_{\pm\pm}$ stands for the probability of obtaining either $^1_1$ or $^{-1}_{-1}$ in both (Alice's and Bob's) measurements. The detailed derivation is presented in Appendix A.

	The probability of equation \ref{eqn:anticorrelation} is essential to compute the performance of the QKD protocol under the effects of the exciton FSS in the QD source. We can use now this expression to compute the total probability of getting correlated measurements in an event in which the bases chosen by Alice and Bob coincide, even if none of those bases correspond to the vertical-horizontal one. Such probability is the addition of the probability when both participants chose $\phi_{1}$ plus the probability when they choose $\phi_{2}$, namely
	
	\begin{equation}\label{eqn:probability_anti}
		\begin{split}
			P_\text{Corr}\,=&\,P_{\phi_{1}}\left[P_{\pm\pm}(\phi_{1},\theta_\text{FSS}) \right] + P_{\phi_{2}}\left[P_{\pm\pm}(\phi_{2},\theta_\text{FSS}) \right],
		\end{split}    
	\end{equation}
	
	where $P_{\phi_{i}}$ for $i=1,2$ is the probability of choosing $\phi_{i}$ as the measurement orientation in a coincident event. Since the election of basis before each measurement is random, then $P_{\phi_{1}} = P_{\phi_{2}} = \frac{1}{2}$. 
	
	%Hence
	%
	%\begin{equation}\label{eqn:probability_anti}
	%	\begin{split}
		%		P_\text{Corr}\,=&\frac{1}{2} \left[P_{\pm\pm}(\phi_{1},\theta_\text{FSS})  + P_{\pm\pm}(\phi_{2},\theta_\text{FSS}) \right].
		%	\end{split}    
	%\end{equation}
	
	In the optical implementation of the original protocol ($\phi_{1}=\pi/8$ and $\phi_{2}=\pi/4$), this total probability turns into
	
	\begin{equation}\label{eqn:probability_anti-specific}
		\begin{split}
			P_\text{Corr}\,=&\frac{5 + 3 C_{\theta_\text{FSS}} }{8} .
		\end{split}    
	\end{equation}
	
	As expected, if the FSS vanishes $P_\text{Corr} = 1$, and the protocol would work optimally in absence of eavesdropping.
	
	For the more general case of the modified protocol, in which the angles defining  $\phi_{1}$ and $\phi_{2}$ are variable, the total probability of correlation for measurements along the coincident orientations reads
	
	\begin{eqnarray}
		%\begin{split}
		P_\text{Corr}\,&=&\,\dfrac{1}{2}\left[ P_{\pm\pm}(\alpha,\theta_\text{FSS}) + P_{\pm\pm}(\alpha+\beta,\theta_\text{FSS}) \right] , \nonumber \\
		&=& \dfrac{1}{2}
		\Big[
		C_{\alpha}^{4} + 
		S_{\alpha}^{4} + 
		2 S_{\alpha}^{2} C_{\alpha}^{2} C_{\theta_\text{FSS}} +  \hspace{1ex} C_{\alpha+\beta}^{4} + 
		S_{\alpha+\beta}^{4} + 
		2 S_{\alpha+\beta}^{2} C_{\alpha+\beta}^{2}C_{\theta_\text{FSS}}
		\Big].
		%\end{split}
		\label{prediction}
	\end{eqnarray}
	
	This expression allows to straightforwardly predict the effects of both, the FSS in the entanglement's source and the directions of the coincident detectors, on the performance of the considered QKD protocol. 
	
	It is important to note that according to this result, while the angles $\alpha$ and $\beta$ are irrelevant in the case $\theta_\text{FSS} = 0$, they become determining on the effectiveness of the protocol when the FSS is not negligible.  
	
	%-----------------  

	Regarding the Bell quantity ($CR$), it is not a fixed value but rather a function that depends on $\alpha$, $\beta$ and $\theta_\text{FSS}$. It can be expressed as 
	
	\begin{equation}\label{eqn:CR_expression}
		\begin{split}
			CR =	 &	C_{\theta_\text{FSS}}
			S_{2(\alpha+\beta)}\left[\dfrac{1}{\sqrt{2}}+S_{2\alpha}\right]	+
			C_{2(\alpha+\beta)}\left[C_{2\alpha}-\dfrac{1}{\sqrt{2}}\right]	+ 	C_{2\alpha}+\dfrac{1}{\sqrt{2}}.
		\end{split}
	\end{equation}

	The details of the derivation are given in Appendix A. 
	
	For the orientations originally proposed by Ekert (in the optical implementation), the equation \ref{eqn:CR_expression} reduces to
	%--------------------------------------------------------------------------------------------------------
	\begin{equation}
		\label{eqn:CHSH_short}
		\left\vert CR \right\vert = \sqrt{2}\left\vert C_{\theta_\text{FSS}}+1\right\vert \leq 2\sqrt{2}.
	\end{equation}
	%--------------------------------------------------------------------------------------------------------
	Thus, although $CR$ is still bounded by $\pm2\sqrt{2}$, it oscillates as $\theta_\text{FSS}$ increases, exhibiting minima ($|CR| = 0$) for  $\theta_\text{FSS} = (2n + 1) \pi$ ($n$ integer).
	%--------------------------------------------------------------------------------------------------------

	%--------------------------------------------------------------------------------------------------------
	
	%--------------------------------------------------------------------------------------------------------
\section{Quantum simulation}\label{sec:4_Computational implementation}
	%--------------------------------------------------------------------------------------------------------
	To test the validity of equation \ref{prediction}, instead of carrying out the delicate and grueling optical experiment, we opt for a quantum computational implementation that simulates the QKD through the considered protocol.
	%------------------------------------------------------
	
	To build a quantum algorithm that emulates the quantum transmission of the key, we first encoded the polarization states into the qubit representation by using the computational basis and rotation gates, as shown in table \ref{tab:table2}.
	
	\begin{table}[H]
       \centering	
		\caption{\label{tab:table2}
			Photon information encoded on quantum computation language}
		\def\arraystretch{2}
			\begin{tabular}{ccc}
				Scheme  &
				Unrotated Basis   &
				Rotated Basis \\
				\hline
				\stackanchor{Photon}{Polarization}  &
				$\big\{ \ket{H}, \ket{V} \big\}$  &   
				$\Big\{ e^{i\phi_{\ell}\hat{Y}/2}\ket{H} , e^{i\phi_{\ell}\hat{Y}/2}\ket{V} \Big\}$ \\
				\stackanchor{Quantum}{Computation}  &
				$\big\{ \ket{0},\ket{1} \big\}$  &
				$\Big\{\hat{R}_{y} (\phi_{\ell})\ket{0},\hat{R}_{y}(\phi_{\ell})\ket{1}\Big\}$ \\
			\end{tabular}
		\end{table}
	
	%------------------------------------------------------
	To reach this goal, we rewrite the dephased singlet state in the computational basis for two qubits, generated by the operator $\hat{Z}\otimes\hat{Z}$, where the parametrized relative phase is added by applying the $R_{z}$-gate over anyone of the two qubits \cite{hernandez2023effects}. Thus, the state of equation \ref{eqn:singlet_state_dephased-linear} reads 
	%------------------------------------------------------
	
	\begin{equation}\label{eqn:singlet_state_computation}
		\vert\psi\rangle\,=\, \dfrac{1}{\sqrt{2}}\left[\vert 0 0 \rangle + e^{-i\theta_\text{FSS}} \vert 1 1 \rangle
		\right].
	\end{equation}

	%------------------------------------------------------
\subsection{Quantum algorithm}
	
	%------------------------------------------------------
	Once the encoding is defined, we focus on creating a quantum circuit that: First, recreates the transmission of a pair of entangled qubits in the state of equation \ref{eqn:singlet_state_computation}. Second, mimics the process of basis selection performed by Alice and Bob, and saves the chosen direction. Third, measures the quantum channels and stores the values registered by each participant.
	
\begin{figure}[H]
		\centering
		\includegraphics[width=0.5\columnwidth]{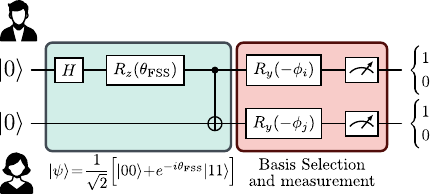}
		\caption{Quantum circuit for the transmission of a bit of information using the modified E91 scheme. In the green stage the dephased Bell state is created and in the pink stage Alice and Bob carry out the measurements along the directions $\phi_{i}$ and $\phi_{j}$, respectively, with $i=0,1,2$ and $j=i+1$. Each measurement may yield either 0 or 1, which are respectively mapped to the -1 or 1 possibilities in the original protocol.}
		\label{fig:quantum_circuit}
	\end{figure}

	The quantum circuit implemented to emulate the generation and transmission of one key-bit is depicted in figure \ref{fig:quantum_circuit}. There, the quantum gates $\boxed{X}$, $\boxed{H}$, $\boxed{R_Z}$,  $\dot{\oplus}$ and $\boxed{R_Y}$, respectively represent the Pauli$_X$, Hadamard, $Z$-rotation, Controlled-$X$ and $Y$-rotation unitary operations. 
	
	The circuit starts by producing the dephased entangled state in terms of $\theta_\text{FSS}$ (green stage). Then introduces the random selection of the direction in which each participant carries out his/her measurement (pink stage). Event by event (associated to each entangled input state), they choose among their corresponding three available directions ($\phi_{i}$ with $i=0,1,2$ for Alice and $\phi_{j}$ with $j=1,2,3$ for Bob), which are defined in terms of the parameters $\alpha$, and $\beta$, according to the orientations shown in figure \ref{fig:ekerts_directions}(b). The implementation of the latter involves a couple of  $\boxed{R_{y}(-\phi_{\ell})}$ rotation gates, that are applied over the quantum channels to emulate the rotation of the polarization detectors, right before the associated measurements. Each measurement may yield either a 0 or a 1, which are correspondingly mapped to -1 or 1, in the language of equation \ref{eqn:anticorrelation}.
	
	Each of these events may become one of the bits in the key-string as long as the bases chosen by Alice and Bob coincide. Hence, to achieve a key with a number of bits long enough, the described process must be repeated a large number of times. 
	
	%This process describes the transition of one key-string, so the key transition implies repeating the quantum circuit stage $N$  times using the same parameters.This process describes the transition of one key-string, so the key transition implies repeating the quantum circuit stage $N$  times using the same parameters.this process must be repeated many times for achieving a string of bits in the key long enough describes the transmission of one key-bit, so the key transition implies repeating the quantum circuit stage $N$  times using the same parameters.
	
	%Then, the quantum circuit stage commences by initializing quantum channels in the $\ket{0}$ state; afterward, the dephased singlet state is created by aplying $\boxed{X}$, $\boxed{H}$, $\boxed{CX}$, and $\boxed{Rz(\theta_\text{FSS})}$ gates as shown in the green block of figure \ref{fig:quantum_circuit}.
	
	%$\alpha$, and $\beta$; and the directions of measurement on the qubits of Alice and Bob, which are pseudo-randomly selected using the set of directions shown in \ref{fig:modified_E91} as we exposed previously, the first three directions are chosen by Alice, and the last three are selected by Bob.
	
	To evaluate the performance of the quantum distribution, we chose the secret key rate ($SKR$) as a convenient metric. This is computed by counting the key-beats successfully transmitted. i.e. key-bits effectively correlated (the same for Alice and Bob in an event in which they chose coincident basis). Readily 
	
	\begin{equation}\label{eqn: Metric}
		{\small
			SKR\,=\,\dfrac{\text{Number of correlated bits in the key-string}}{\text{Number of events measured in coincident basis}}.}
	\end{equation}
	
	A complementary metric is the so-called quantum bit error rate ($QBER$), that oppositely to the $SKR$, focus on the number of bits transmitted with error (anticorrelated or unmeasured), in an event in which Alice and Bob chose coincident bases. In this case in which eavesdropping and leaking is not considered, $QBER = 1-SKR$.

	%------------------------------------------------------
	
	%------------------------------------------------------
	%what allows us to evaluate the performance just by counting the successfully transmitted states \cite{hernandez2023effects}. 
	%------------------------------------------------------
	
	%------------------------------------------------------
	In the final stage of the computational implementation, the $SKR$ value is registered as a function of the parameters $\alpha$, $\beta$ and $\theta_\text{FSS}$. 
	%------------------------------------------------------
	%This is repeated until completing the whole values of the parameters.
	%------------------------------------------------------
	
	%------------------------------------------------------

		%--------------------------------------------------------------------------------------------------------
\section{Results}\label{sec:5_Result}

\subsection{Simulations and Discussion}
	
	We execute the algorithm described in the previous section, to replicate the modified version of E91 protocol by means of the IBM's Qiskit Aer simulator \cite{qiskit2024}.
	%--------------------------------------------------------------------------------------------------------
	We opted for executing it in a quantum simulator instead of an actual quantum processor to avoid the effects of noise, which we expect to incorporate into the model in a further work.
	
		For the simulations, the parameters $\alpha$ and $\beta$  ($\theta_{FSS}$) were varied within the interval $\left[0,\pi\right]$ ($\left[0,2\pi\right]$). Nonetheless, only multiples of $\pi/8$ were considered for $\alpha$. 
	%while for  we set a scan in steps of $\pi/4$ in the same interval.
	
	For each execution of the protocol, associated to a set of $\theta_\text{FSS}$, $\alpha$, and $\beta$ values, a total of $5\times10^{4}$ events (entangled input states) were used to reduce the error margin in the $SKR$. Overall, we run $10^{4}$ executions of the protocol. 
	
\begin{figure}[H]
		\centering
		\subfloat[$\alpha=\pi/8$. $R^{2}=0.997$\label{fig:appx_plot_perf_pi_4}]{\includegraphics[width=.5 \columnwidth]{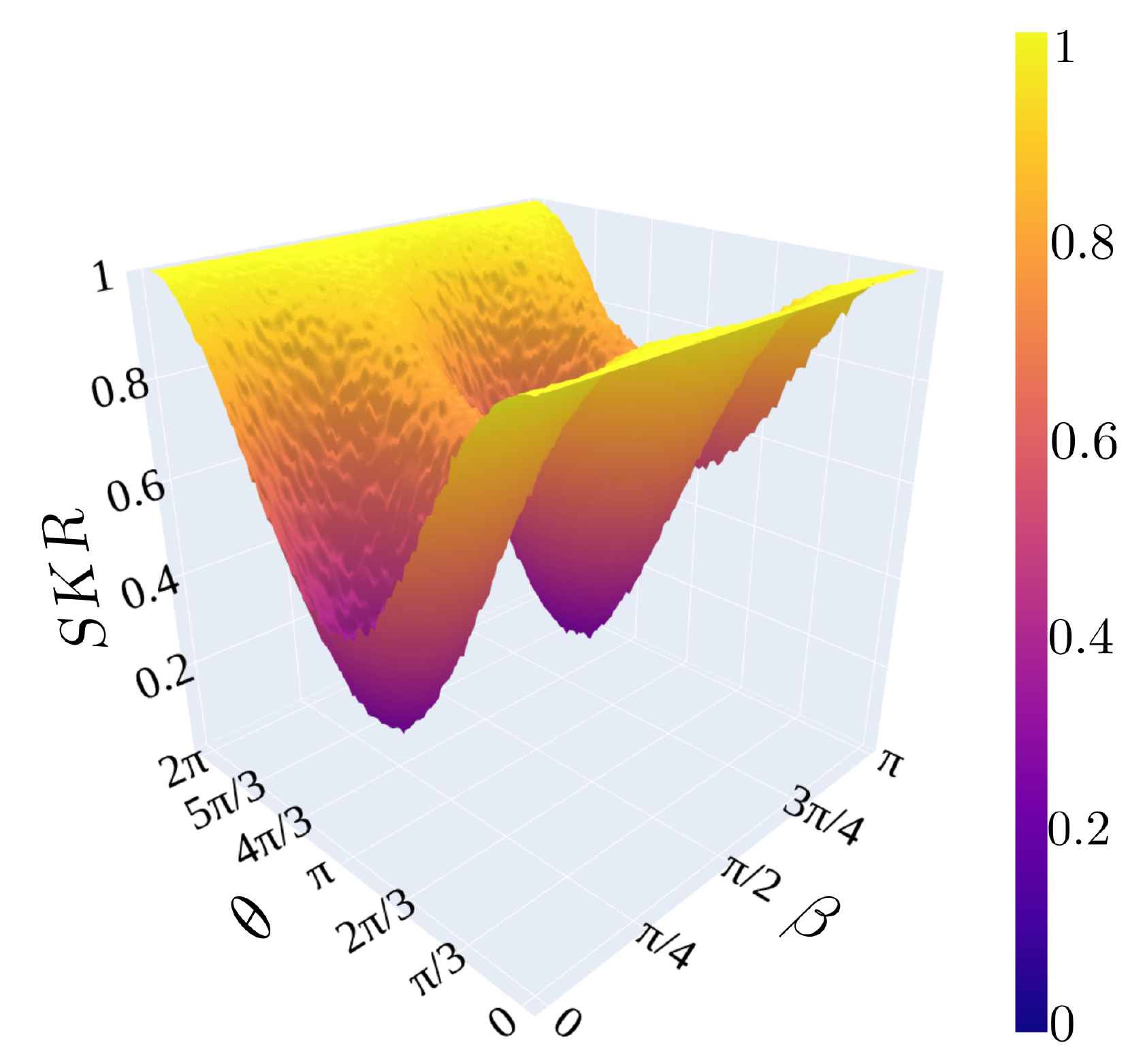}}
		\hfill
		\subfloat[$\alpha=\pi/4$. $R^{2}=0.994$\label{fig:appx_plot_perf_pi_2}]{\includegraphics[width=.5 \columnwidth]{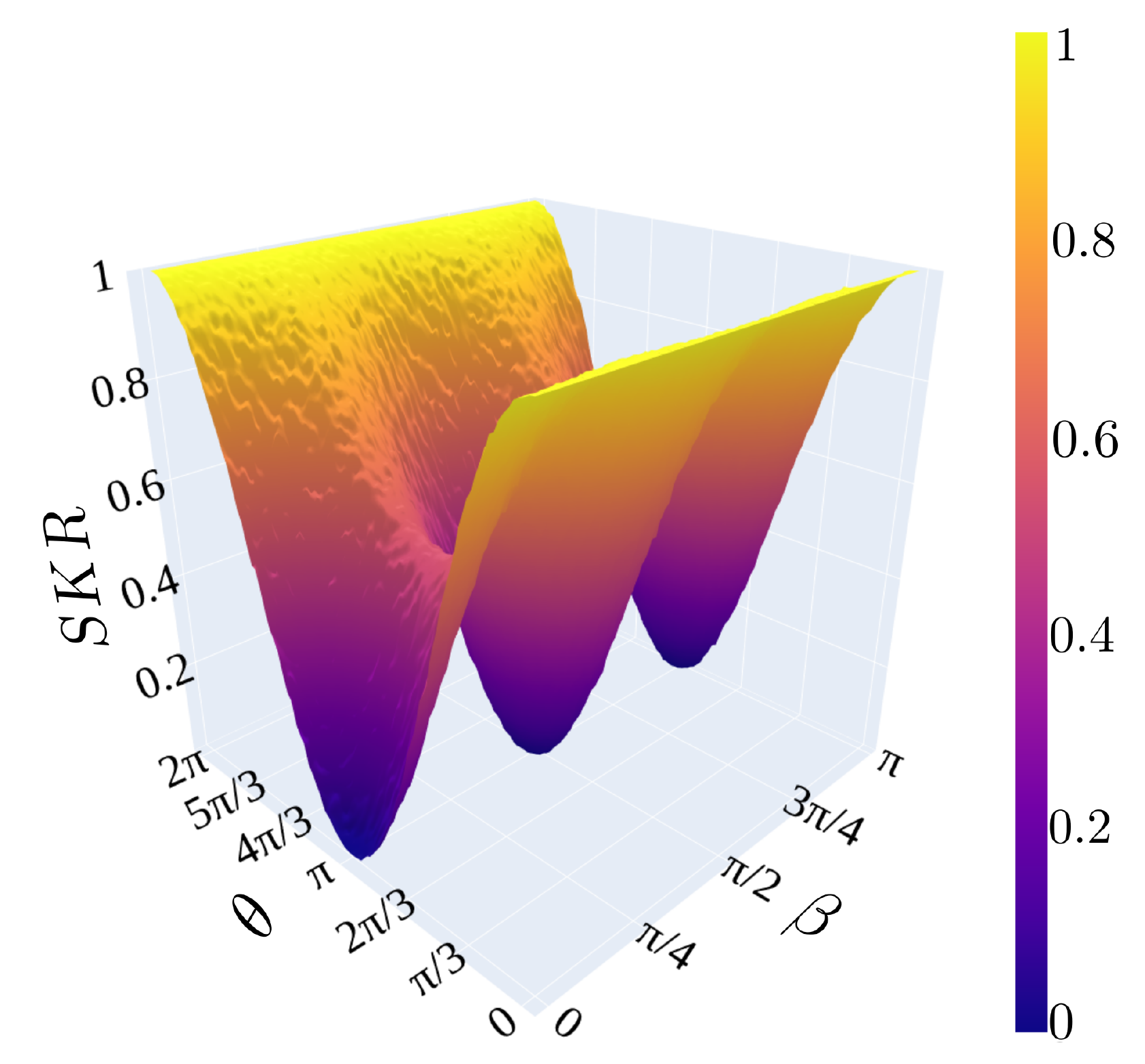}}
		\\
		\subfloat[$\alpha=3\pi/8$. $R^{2}=0.997$\label{fig:appx_plot_perf_3_pi_4} ]{\includegraphics[width=.5  \columnwidth]{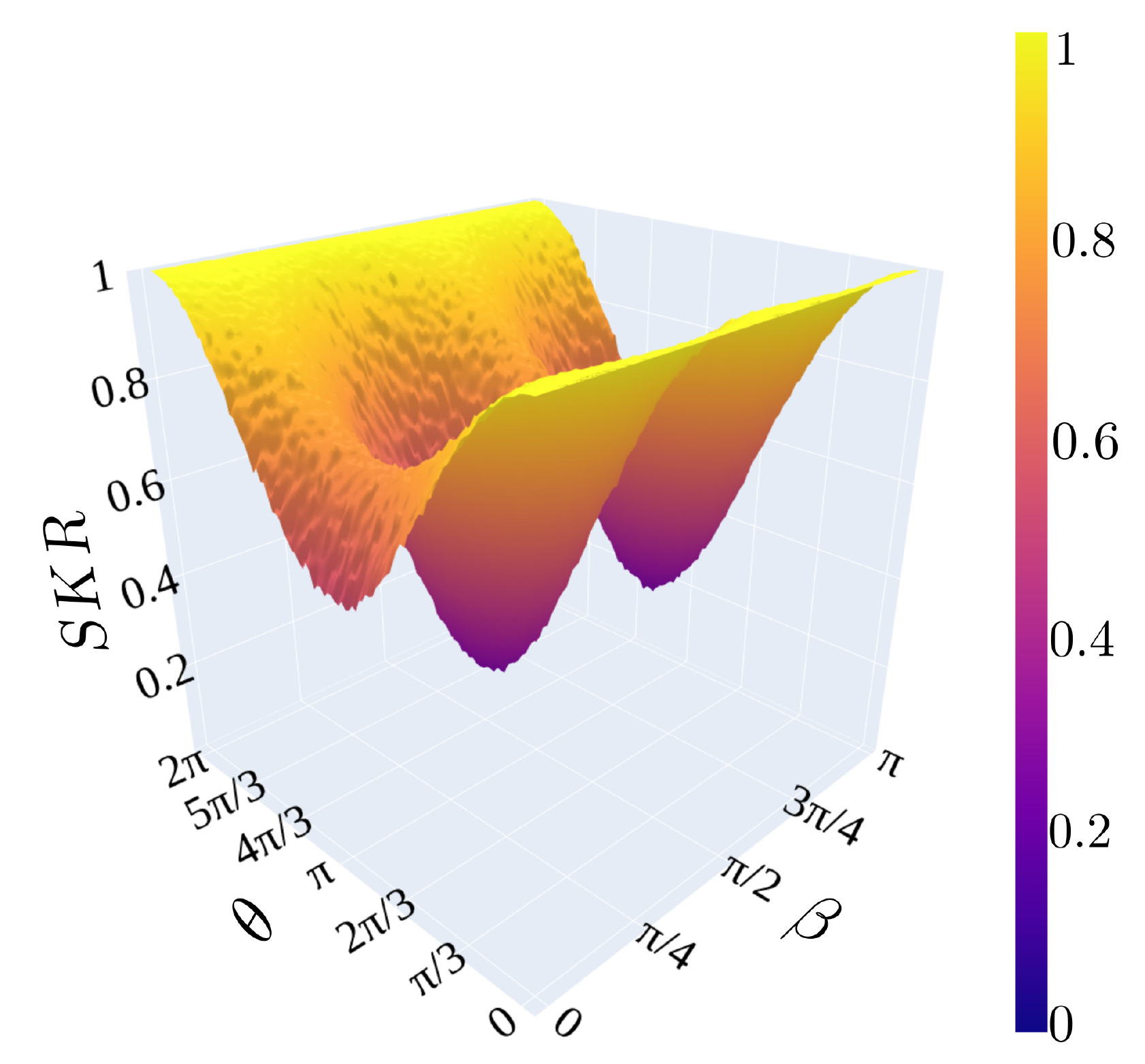}}
		\hfill
		\subfloat[$\alpha=\pi/2$.  $R^{2}=0.917$\label{fig:appx_plot_perf_pi}]{\includegraphics[width=.5 \columnwidth]{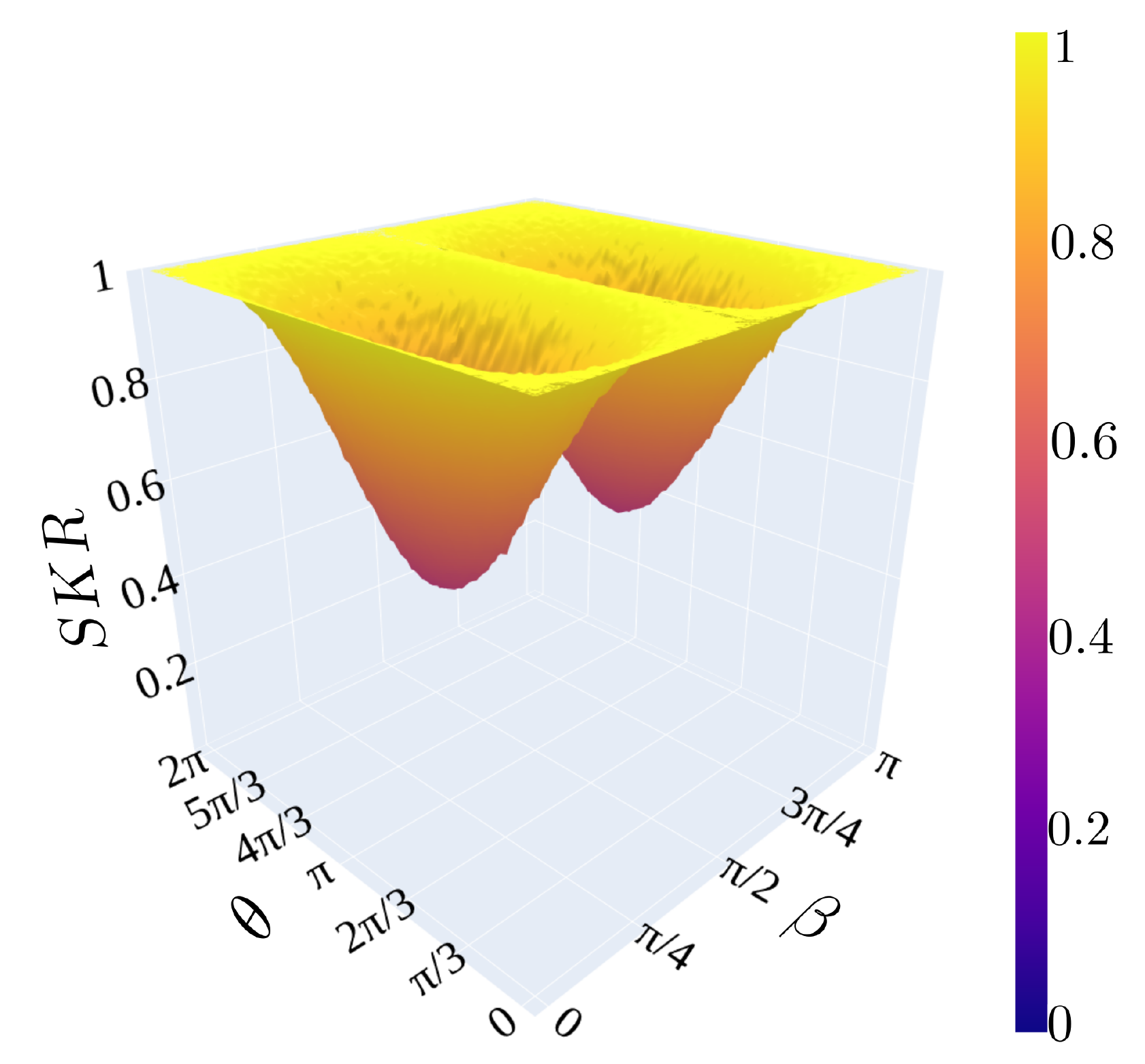}}
		
		\caption{SKR for $\alpha=\ell\dfrac{\pi}{8}$ with $\ell=1,2,3,4$. The correlation coefficient $R^2$ in each panel, indicates the coincidence between the regression of  the corresponding data and the analytical expression (equation \ref{prediction}).}		
		\label{fig:alpha_plots}
	\end{figure}

	Figure \ref{fig:alpha_plots} shows the quantum computing simulations for $SKR$, obtained for $\alpha=\ell\pi/8$ with $\ell=1,2,3,4$, as functions of $\theta_{FSS}$ and $\beta$.
	
	A correlation index $R^{2} \ge 0.92$ between the simulation data and the expression in equation \ref{prediction} was obtained for all the shown cases, representing a quite satisfactory match between the computational implementation and the analytical results, which validates our model. 
	
	The surface plots show a strong dependence of $SKR$ on $\theta_\text{FSS}$, for most values of $\beta$, presenting the minimum performance in $\theta_\text{FSS}=\pi$. The only exceptions to this behavior are observed in the cases $\alpha = n \pi/2$ with $n$ integer, in which $SKR$ is independent on $\theta_\text{FSS}$ for $\beta = n \pi/2$. This is a trivial an useless scenario for which both polarizers are along the same direction, and that coincides with the orientation defined by the computational basis. In fact, any case in which $\beta$ is a multiple of $\pi$, independently on the value of $\alpha$, corresponds to a condition under which the protocol cannot be securely applied, because the randomness in the election of basis for measurements would be lost, and interception of Alice's or Bob's photons may result in exposition of the key.  
	
	It can be seen in figure \ref{fig:alpha_plots}, how $SKR$ oscillates on $\beta$ with a period of $\pi/2$, and that the position and depth of the minima depend on $\alpha$. Transversal cuts at different values of $\beta$ have alike features in the considered interval. It initiates with $SKR=1$, indicating an optimal performance of the protocol. Then, it decays until a minimum at $\theta_\text{FSS}=\pi$, where it starts increasing again until recovering ideal operation of the protocol at $\theta_\text{FSS}=2\pi$.  
	
	There are two scenarios in which the minimum $SKR$ reach extreme values. $\alpha = n \pi/2 $  and $\alpha = (2n+1)\pi/4$ ($n=0,1,2,3...$). 
	
	In the former case, the minimum $SKR$ value is $0.5$, indicating total randomness in the correlation of Alice's and Bob's measurements. In that configuration, the minima are located in $\beta = (2n+1)\pi/4$.  Something expected, because orthogonality of the detection polarizers leads to minimum correlation \cite{grynberg2010introduction}.
	
	In the latter case, the lowest value for $SKR$ is $0$, which interestingly implies that in such configuration, for $\theta_\text{FSS}=\pi$, the intended correlation in Alice's and Bob's measurements, turns into perfect anticorrelation.
	
	The data from the simulations also allow to validate the model regarding the predictions on the $\left\vert CR\right\vert$ quantity. Hence, we calculated such a quantity in terms of $\theta_\text{FSS}$ to compare the results with the corresponding derived expression in equation \ref{eqn:CR_expression}. 
	%--------------------------------------------------------------------------------------------------------
	
\begin{figure}[H]
		\centering
		\includegraphics[width=.5\columnwidth]{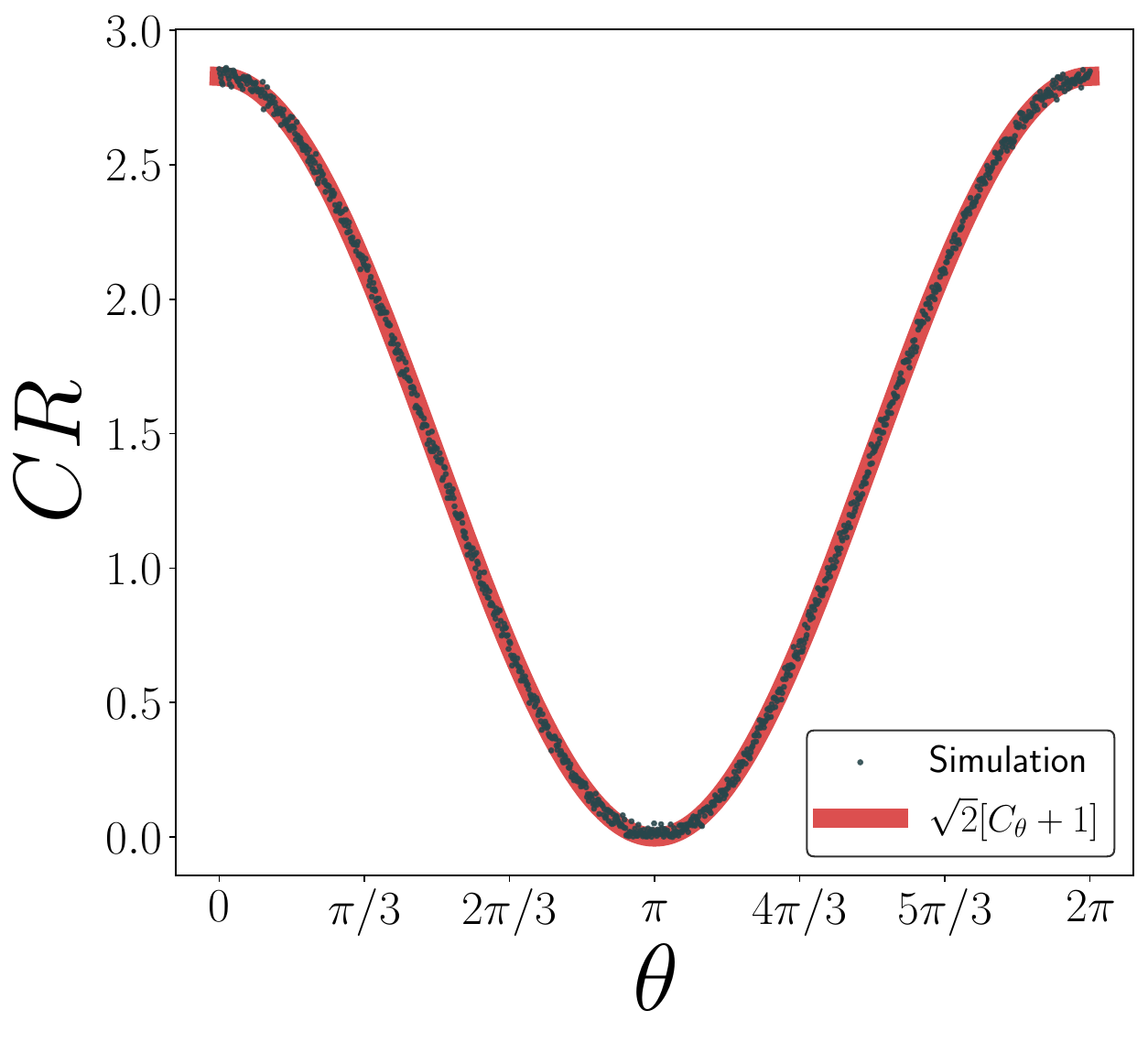}
		\caption{Derived $\left\vert CR \right\vert$, from equation \ref{eqn:CR_expression}, as a function of $\theta_\text{FSS}$ (red solid line) and the corresponding quantity calculated for 1000 executions with different values of $\theta_\text{FSS}$ (black dots). $\alpha=\beta=\pi/8$ were used. The correlation index between the analytical prediction and the simulations is $R^2 = 0.99$.}
		\label{fig:chsh_plot}
	\end{figure}

	Figure \ref{fig:chsh_plot} shows the correspondence between the analytical expression and the data from the simulations for the case $\alpha=\pi/8$ and $\beta=\pi/8$, i.e. the same orientations proposed originally by Ekert to maximize the violation of the Bell's inequality CHSH. The protocol was executed $10^{3}$ times, with $5\times10^{4}$ events (entangled input states) per execution. A high correlation between the predicted $CR$ function and the results from the quantum computing implementation was found,  with an index of $R^{2} = 0.99$. 
	%--------------------------------------------------------------------------------------------------------

	The oscillations of $CR$ on $\theta_\text{FSS}$ imply that computing such a quantity to prove security in  the key transmission, may trigger spurious alerts because $ |CR| < 2$ could happen even in absence of eavesdropping, as long as the FSS in the QD source is not negligible. This suggests that in presence of non-vanishing $S$ for the entanglement source, alternative mechanisms for verifying the security of the transmission should be devised. An option, if the photon leaking rate $LKR$ for transmission and detection is well characterized (in our case it is taken as zero), would be to test 
	
	\begin{equation}
		\label{eqn:new-test}
		SKR + QBER + LKR = 1 ,
	\end{equation}   
	
	which should be fulfilled if no eavesdropping were present. 
	
	Summarizing, our analytical and computational results prove that the performance of the QKD protocol E91 depends substantially on the dephasing in the input entangles state. For a wide range of $\theta_\text{FSS}$ values, such a dephasing strongly diminishes the ability of the protocol for transmitting reliable keys.
	
	Besides effective reduction of either the FSS or the exciton lifetime in the QD used to produce the entangled states, a possibility for mitigating the unfavorable effects of the dephasing is to adjust the detection angles in the protocol implementation. According to figure \ref{fig:alpha_plots}, $\alpha$ and $\beta$ could be tuned to rise the minimum $SKR$, extending the range of $\theta_\text{FSS}$ values along which $QBER$ would remain within the acceptable regime established by the Shannon limit ($<0.11$) \cite{Shannon-limit,Shannon-limit-2021}.      
	
	It is worth remarking that although these results were obtained considering the effects of the FSS on the entangled state produced by a QD source, they are applicable in situations where other sources of dephasing can be considered, e.g. recapture, valence band mixing and exciton-spin flipping \cite{review-entanglement-source}.  
	
	Finally, we would like to highlight that the validation of the analytical model by means of the quantum computational approach resulted much cheaper and faster than what the experimental counterpart would have been. These simulations were carried out in an average commercial-type desktop machine and the computing times were at the order of days.    
	
\subsection{Protocol BBM92 as a limit}
	
	The entanglement-based QKD protocol known as BBM92 \cite{bennett1992quantum}, can be obtained as a particular case of the modified E91 scheme described in section \ref{sec:3_E91 Protocol}, by setting $\alpha = 0$ and $\beta = \pi/4$ (see figure \ref{fig:ekerts_directions}(b)). It was devised by C. Bennett, G. Brassard and N. Mermin to securely transmit a key by using only two directions, corresponding to horizontal-vertical ($\{\ket{H},\ket{V}\}$) and diagonal-antidiagonal ($\{\ket{D},\ket{A}\}$) polarizations. 
	
	The BBM92 protocol has been chosen for QKD experimental implementations because of its simplicity, since it does not include Bell inequality measurements \cite{experimental-BBM92}. 
	
	Inserting those specific values ($0$ and $\pi/4$) into equation \ref{prediction}, the corresponding correlation probability becomes
	
	\begin{equation}
		P_{Corr}= \dfrac{3+C_{\theta_{FSS}}}{4},
	\end{equation}
	
	which coincides with the previously reported expression for that protocol, in reference \cite{hernandez2023effects}.
	
	Nevertheless, the formulation here presented unlocks the possibility of enhancing the SKR for a given value of $\theta$, by means of the observed $\beta$-dependence.  Such correlation probability becomes
	
	\begin{equation}
		P_{Corr} = \dfrac{1}{2} \left[ 1 + C_{\beta}^4 + S_{\beta}^4 + 2S_{\beta}^{2}C_{\beta}^{2}\, C_{\theta_{FSS}} \right],
	\end{equation}
	
	which reveals a mechanism to improve the efficacy of the BBM92 protocol under dephasing of the input state, by uplifting its minimum SKR (see figure \ref{fig:alpha_plots}(d)).

%	By setting our model to $\alpha =0$, $ \beta=\pi/4$ it is possible to compute the SKR computed for BBM92
%	
%	\begin{equation}
%		\begin{split}
%			P_{Corr}&= \dfrac{1}{2} \left[ 1 + \left(C_{\frac{\pi}{4}}\right)^4 + \left(S_{\frac{\pi}{4}}\right)^4 + 2\left(S_{\frac{\pi}{4}}C_{\frac{\pi}{4}}\right)^2 C_{\theta_{FSS}} \right]
%			\\
%			&= \dfrac{1}{2} \left[ 1 + \dfrac{1}{2} + \dfrac{C_{\theta_{FSS}}}{2} \right]
%			\\
%			&= \dfrac{3 + C_{\theta_{FSS}}}{4}.
%		\end{split}
%	\end{equation}
%	
%	Our group previously presented this result, which confirms our model and shows how BBM92 can be a specific case of E91 \cite{hernandez2023effects}.  
	
	%--------------------------------------------------------------------------------------------------------
	
	%--------------------------------------------------------------------------------------------------------
\section{Conclusions}\label{sec:7_Conclusions}
	We studied the performance of the QKD protocol E91 under the effects of dephased polarization-entangled states generated by radiative cascade in a QD source. We also investigated the influence of varying the orientation of the measurement polarizers by parameterizing the directions in which there can be coincidence of the analyzers.
	%---------------------------------------------------
	
	%---------------------------------------------------
	We derived explicit expressions for the performance of the quantum distribution and for the quantity used to carry out the CHSH-type Bell test in the stage of security verification, as functions of the phase of the entangled state and of the varying angles in the protocol realization. Those analytical expressions were satisfactorily validated by multiple simulations from a quantum computing implementation of the protocol, executed in the IBM's Qiskit Aer simulator.
	%---------------------------------------------------
	
	%---------------------------------------------------
	According to our results, the secret key rate for the transmission is substantially affected by the dephasing in the entangled state. Under some conditions, the value of this rate may be as low as 0.5, which corresponds to a scenario where the protocol becomes completely ineffective. 
	
	In turn, the parameter for the Bell test also oscillates on the magnitude of the dephasing, reaching values in which the CHSH inequality is not violated. This may lead to false positives in the stage of eavesdropping detection.   
	
	The observed dependence of the secret key rate on the varying angles of the measurement polarizers, suggest a mechanism to remediate the adverse effects generated on the reliability of the key transmission by the exciton fine structure splitting of the quantum dot used to produce the entangled states. 
	
	These findings, that are extendable to the BBM92 protocol since it can be seen as a particular case within the presented formulation, provide valuable insight on standing challenges and potential solutions toward the large scale use of novel sources of entanglement in emerging technologies like quantum communication and quantum cryptography. 
	
	From a broader point of view, this work represents a vivid example of how currently available quantum computation tools are useful for simulating physical systems and processes whose experimental realization may turn arduous.

\section*{Acknowledgments}	
		The authors acknowledge support from the Colombian SGR through project BPIN2021000100191 and from the Research Division of UPTC through Project No. SGI-3378.

	\appendix
	%--------------------------------------------------------------------------------------------------------
\section{Detailed derivations}
	%------------------------------------------------------------------------------------------------
\subsection{Calculation of the correlation probability\label{apx:correlation_calculation}}
	%------------------------------------------------------------------------------------------------
	Let's suppose that Alice and Bob respectively set their analyzers in the directions $\phi_{a}$ and $\phi_{b}$. The horizontal and vertical polarization eigenvalues ($\ket{H}$ and $\ket{V}$) are given by 
	
	\begin{eqnarray}
		\ket{H} &= C_{\phi_{\ell}}\ket{+_{\phi_{\ell}}} + S_{\phi_{\ell}}\ket{-_{\phi_{\ell}}}, \nonumber \\
		\ket{V} &= -S_{\phi_{\ell}}\ket{+_{\phi_{\ell}}} + C_{\phi_{\ell}}\ket{-_{\phi_{\ell}}},
	\end{eqnarray}
	
	in terms of the corresponding Alice's and Bob's polarization eigenvalues 	$\ket{\pm_{\phi_{\ell}}}$, with $\ell=a,b$
	\cite{grynberg2010introduction}. 
	
	Thus, the initial state of equation \ref{eqn:singlet_state_dephased-linear}  
	
	\begin{equation}\label{eqn:singlet_state_dephased-linear-appendix}
		\vert\psi\rangle\,=\, \dfrac{1}{\sqrt{2}}\left[\vert H_{XX} H_{X} \rangle + e^{-i\theta_\text{FSS}} \vert V_{XX} V_{X} \rangle
		\right].
	\end{equation}
	
	can be rewritten in the Alice's and Bob's basis, according to  
	
	\begin{equation}\label{eqn:rotated bell state}
		\begin{split}
			\ket{\psi} =\dfrac{1}{\sqrt{2}}
			\Big[
			&\ket{+_{\phi_{a}}+_{\phi_{b}}}
			\left(C_{\phi_{a}}C_{\phi_{b}} + e^{-i\theta_{FSS}}S_{\phi_{a}}S_{\phi_{b}}\right) +
			\ket{-_{\phi_{a}}-_{\phi_{b}}}
			\left(S_{\phi_{a}}S_{\phi_{b}} + e^{-i\theta_{FSS}}C_{\phi_{a}}C_{\phi_{b}}\right) 
			\\
			+ & \ket{+_{\phi_{a}}-_{\phi_{b}}}
			\left(C_{\phi_{a}}S_{\phi_{b}} - e^{-i\theta_{FSS}}S_{\phi_{a}}C_{\phi_{b}}\right) +
			\ket{-_{\phi_{a}}+_{\phi_{b}}}
			\left(S_{\phi_{a}}C_{\phi_{b}} - e^{-i\theta_{FSS}}C_{\phi_{a}}S_{\phi_{b}}\right)
			\Big].
		\end{split}
	\end{equation}
	
	Because of orthogonality of the Alice's and Bob's bases, the probabilities of having correlated measurements, $P_{++}$ and $P_{--}$, can now be straightforwardly obtained through the corresponding projections. Then
	
	\begin{equation}
	\label{appendix-6}	
	\begin{split}
		P_{++}	&=\Big|\langle+_{\phi_{a}},+_{\phi_{b}}\vert\psi\rangle\Big|^{2} ,
		\\
		&=\dfrac{1}{2}\Big|C_{\phi_{a}}C_{\phi_{b}}+e^{-i\theta_{FSS}}S_{\phi_{a}}S_{\phi_{b}}\Big|^{2} ,
		\\
			&=\dfrac{1}{2}\left[C_{\phi_{a}}^{2}C_{\phi_{b}}^{2}+S_{\phi_{a}}^{2}S_{\phi_{b}}^{2}+2S_{\phi_{a}}C_{\phi_{a}}S_{\phi_{b}}C_{\phi_{b}}C_{\theta_{FSS}}\right].
	\end{split}
\end{equation}	

Since the transmitted key is composed of measurements done in matching directions, we take $\phi_{a}=\phi_{b}=\phi$ and equation \ref{appendix-6} becomes 

\begin{equation}
	\label{appendix-7}
	P_{++}	= \dfrac{1}{2} \left[C_{\phi}^{4} + S_{\phi}^{4} + 2S_{\phi}^{2}C_{\phi}^{2}C_{\theta_{FSS}}\right],
\end{equation}

inline with equation \ref{eqn:anticorrelation}.  

The calculation for $P_{--}$ is analogous and yields the same result.

The probability of having anticorrelated measurements, $P_{- +}$ (which is the same as $P_{+ -}$), is

	\begin{equation}
		\label{appendix-8}
		\begin{split}
			P_{+-}	&=\Big|\langle\pm_{\phi_{a}},\mp_{\phi_{b}}\vert\psi\rangle\Big|^{2} ,
			\\
			&=\dfrac{1}{2}\Big|C_{\phi_{a}}S_{\phi_{b}}-e^{-i\theta_{FSS}}S_{\phi_{a}}C_{\phi_{b}}\Big|^{2} ,
			\\
			&=\dfrac{1}{2}\left[C_{\phi_{a}}^{2}S_{\phi_{b}}^{2}+S_{\phi_{a}}^{2}C_{\phi_{b}}^{2} - 2S_{\phi_{a}}C_{\phi_{a}}S_{\phi_{b}}C_{\phi_{b}}C_{\theta_{FSS}}\right] .
		\end{split}
	\end{equation}
	%--------------------------------------------------------
	
\subsection{Derivation of the angle-dependent $CR$ quantity}\label{appx:computational result}
	%--------------------------------------------------------
	
	Now, we can use the probabilities of correlated and anticorrelated measurements to obtain the correlation coefficients $E_{ab} \equiv E(\phi_{a},\phi_{b},\theta_{FSS})$, where $\phi_{a}$ ($\phi_{b}$) is the direction selected by Alice (Bob). Those coefficients are given by \cite{clauser1969proposed}
	
	\begin{equation}
		\label{eqn:coefficient_correlation-1}
			E_{ab} = P_{++} + P_{--} - P_{+-} - P_{-+}.
	\end{equation}
	
	Thus, inserting equations \ref{appendix-6} and \ref{appendix-8} into equation \ref{eqn:coefficient_correlation-1}, the correlation coefficients turn into  
	
	\begin{equation}
		\label{eqn:coefficient_correlation-2}
		\begin{split}
			E_{ab} = &
			C_{\phi_{a}}^{2} C_{\phi_{b}}^{2} +
			S_{\phi_{a}}^{2} S_{\phi_{b}}^{2} +
			2S_{\phi_{a}}C_{\phi_{a}} S_{\phi_{b}}C_{\phi_{b}} C_{\theta_{FSS}} \\ & -
			C_{\phi_{a}}^{2} S_{\phi_{b}}^{2} -
			S_{\phi_{a}}^{2} C_{\phi_{b}}^{2} +
			2S_{\phi_{a}}C_{\phi_{a}} S_{\phi_{b}}C_{\phi_{b}}  C_{\theta_{FSS}},
			\\
			= &
			C_{\phi_{a}}^{2}\left[ C_{\phi_{b}}^{2} - S_{\phi_{b}}^{2} \right] -
			S_{\phi_{a}}^{2}\left[ C_{\phi_{b}}^{2} - S_{\phi_{b}}^{2} \right] + 
			4S_{\phi_{a}}C_{\phi_{a}} S_{\phi_{b}}C_{\phi_{b}}  C_{\theta_{FSS}},
			\\
			 = &
			\left[ C_{\phi_{a}}^{2} - S_{\phi_{a}}^{2} \right]C_{2\phi_{b}} + S_{2\phi_{a}}S_{2\phi_{b}}C_{\theta_{FSS}},
			\\
			= &
			C_{2\phi_{a}} C_{2\phi_{b}} + S_{2\phi_{a}}S_{2\phi_{b}}C_{\theta_{FSS}}.		
		\end{split}
	\end{equation}
	%--------------------------------------------------------------------------------------------------------
	
	%--------------------------------------------------------------------------------------------------------
	Hence, the $CR$ quantity to test the CHSH inequality is given by
	%--------------------------------------------------------------------------------------------------------
	\begin{equation}\label{eqn:CR_expression-appendix}
		\begin{split}
			CR = & E_{01} + E_{23} - E_{03} + E_{21} ,
			\\
			 = &  	C_{2\phi_{0}} C_{2\phi_{1}} + S_{2\phi_{0}}S_{2\phi_{1}}C_{\theta_{FSS}} +
			C_{2\phi_{2}} C_{2\phi_{3}} + S_{2\phi_{2}}S_{2\phi_{3}}C_{\theta_{FSS}} \\ & -
			C_{2\phi_{0}} C_{2\phi_{3}} - S_{2\phi_{0}}S_{2\phi_{3}}C_{\theta_{FSS}} +
			C_{2\phi_{2}} C_{2\phi_{1}} + S_{2\phi_{2}}S_{2\phi_{1}}C_{\theta_{FSS}},		
			\\
			= &	C_{\theta_{FSS}}
			\Big[
			S_{2\phi_{0}}\left(S_{2\alpha} -S_{2\phi_{3}}\right) +
			S_{2(\alpha+\beta)}\left(S_{2\phi_{3}}+S_{2\alpha}\right)
			\Big]	\\ & +
			C_{2\phi_{0}}\left(C_{2\alpha}-C_{2\phi_{3}}\right)	+
			C_{2(\alpha+\beta)}\left(C_{2\phi_{3}} + C_{2\alpha}\right),
		\end{split}
	\end{equation}
	%---------------------------------------------------------
	in which subsitution of the values $\phi_{0}=0$ and $\phi_{3}=3\pi/8$ leads to 
	
	%---------------------------------------------------------
	\begin{equation}
		\begin{split}
			CR =	 &	C_{\theta_{FSS}}
			S_{2(\alpha+\beta)}\left[\dfrac{1}{\sqrt{2}}+S_{2\alpha}\right]	+
			C_{2(\alpha+\beta)}\left[C_{2\alpha}-\dfrac{1}{\sqrt{2}}\right]	+ C_{2\alpha}+\dfrac{1}{\sqrt{2}},
		\end{split}
	\end{equation}
	
	that corresponds to equation \ref{eqn:CR_expression}.
	%--------------------------------------------------------------------------------------------------------
	
%\bibliographystyle{abbrvnat}
%\bibliographystyle{elsarticle-num}
\bibliography{FSS-effects-on-E91-QKD-CPC-vfv.bib}% Produces the bibliography via BibTeX.
	
\end{document}